\documentclass[twocolumn,prl,showpacs]{revtex4}
\usepackage{graphicx}
\usepackage{dcolumn}
\usepackage{bm}
\usepackage{color}

\begin{document}
\title{Spin-triplet f-wave pairing due to three-site cyclic-exchange ferromagnetic interactions}
\author{S.~Nishimoto}
\affiliation{Leibniz-Institut f\"ur Festk\"orper- und Werkstoffforschung 
Dresden, D-01171 Dresden, Germany}
\author{Y.~Ohta}
\affiliation{Department of Physics, Chiba University, Chiba 263-8522, Japan}

\date{\today}

\begin{abstract}
Ferromagnetiam and superconductivity in a two-dimensional triangular-lattice 
Hubbard model are studied using the density-matrix renormalization group method. 
We propose a mechanism of the {\it f}-wave spin-triplet pairing derived 
from the three-site cyclic-exchange ferromagnetic interactions. We point out that 
a triangular network of hopping integrals, which is required for the three-site 
cyclic hopping processes, is contained in the (possibly) spin-triplet superconducting 
systems, such as Bechgaard salts (TMTSF)$_2$X, cobalt oxide 
Na$_{0.35}$CoO$_2$$\cdot$$1.3$H$_2$O, and layered perovskite Sr$_2$RuO$_4$.

\end{abstract}
\maketitle

Symmetry of Cooper pairs give an invaluable information of 
our elucidating a superconducting mechanism. Both spin and 
angular momentum are good quantum numbers for describing 
the Cooper pairs. In the spin-triplet channel, the total spin 
is one and the pairing symmetry is of {\it p}- or {\it f}-wave. 
The long-standing quest for a new example of spin-triplet 
superconductivity has been pursued since the spin-triplet 
superfluid $^3$He was discovered~\cite{Leggett75}. 
Recently, three experimental candidates have been hand-running 
recognized to be likely spin-triplet superconductivity (see below). 
They provide an excellent chance for us to have productive 
insights into the nature of spin-triplet pairing, which is less 
understood than that of the spin-singlet pairing. Spin-triplet 
superconductivity is thus one of the hottest topics in the 
field of strongly-correlated electron systems.

From the theoretical aspect, spin-triplet Cooper pairs are considered 
to be formed simply by a ferromagnetic interaction~\cite{Fay80}. 
The realization of such pairs in a system of electrons interacting 
via spin-independent Coulomb interaction, namely the Hubbard model, 
is still a significant challenge. The Nagaoka-Thouless~\cite{Nagaoka66,Thouless65} 
and flat-band~\cite{Mielke91} mechanisms are well known as the origin 
of ferromagnetism. However, both of them are unsuitable for explaining 
the spin-triplet superconductivity since they lead not to Cooper pairs 
but to saturated magnetization. Then one may turn his eyes to 
the three-site cyclic-exchange ferromagnetic interaction which is only 
the remaining mechanism in the simple Hubbard model~\cite{Tasaki98}. 
It brings a couple of electrons to be locally ferromagnetic, so that 
applying the mechanism to the spin-triplet superconductivity seems 
to be a natural extension. In this letter, we propose a mechanism 
of the spin-triplet superconductivity derived from the three-site 
cyclic-exchange ferromagnetic interaction. 

We are also highly motivated by the fact that all of the three 
experimental candidates contain a triangular network of hopping 
integrals which is required for the three-site cyclic-exchange mechanism.
One of the candidates is Bechgaard salts (TMTSF)$_2$X~\cite{Bechgaard80}. 
The crystal structure consists of well-separated sheets containing 
one-dimensional (1D) TMTSF stacks along the {\it a}-axis. The sheets 
are in the {\it ab}-plane and the hopping integrals along the {\it b}-axis 
are about $10-20\%$ of those along the {\it a}-axis~\cite{Ducasse86}. 
The unique structure of the hopping integrals can be regarded 
as an anisotropic triangular lattice. In experiment~\cite{Lee06}, it has 
been widely concluded that the superconducting state is of the spin-triplet 
pairing: the spin susceptibility, from the NMR Knight shift measurement, 
remains unchanged through the superconducting transition ($T_{\rm c}$) 
and the upper critical magnetic field exceeds substantially 
the Pauli limit in both {\it a} and {\it b} directions. 
Also, the enhancement of the spin-lattice relaxation ratio just below 
$T_{\rm c}$, indicating the presence of line nodes in the 
superconducting order parameter, was found~\cite{Takigawa87}. 
In theory~\cite{Kuroki06-1}, we seem to have reached a consensus 
on the spin-triplet pairing; nevertheless, the pairing symmetry is still 
open issue~\cite{Abrikosov83,Hasegawa87,Lebed99,Kuroki01,Nickel05,Tanaka04-1,Fuseya05}.

Another experimental candidate is cobalt oxide Na$_{0.35}$CoO$_2$$\cdot$$1.3$H$_2$O, 
which shows a superconducting transition at $T_{\rm c} \approx 5$ K~\cite{Takada03}. 
The conductive {\it ab} planes consist of edge-sharing CoO$_6$ octahedra 
and each plane is strongly separated by Na$^+$ ions and H$_2$O molecules 
along {\it c} axis.  The Co ions form a triangular lattice, and the system may 
be regarded as a two-dimensional (2D) triangular lattice doped with $35\%$ electrons. 
Some experiments have suggested the possibility of an unconventional 
superconductivity with line nodes in the gap 
function~\cite{Kato06,Sakurai03,Fujimoto04,Higemoto04}. 
In particular, spin-triplet {\it p}- or {\it f}-wave pairing state has been 
argued based on an invariant bahavior of the {\it ab}-plane Knight shift 
across $T_{\rm c}$ and a $T^3$-dependence of $1/T_1$ below $T_{\rm c}$~\cite{Kato06}. 
On the other hand, a suppression of Knight shift across $T_{\rm c}$ for 
both magnetic field parallel and perpendicular to the {\it ab}-plane suggesting 
a spin-singlet pairing state has also been reported~\cite{Kobayashi08}. 
It is supported by an angle-resolved photoemission study, which have observed 
only hole-like Fermi surface around $\Gamma$ point~\cite{Shimojima06}. 
As described above, experimental decision of the pairing symmetry is still 
controversial. Similarly, theoretical studies are not in agreement: 
the {\it d}+{\it id}- (or {\it p}-)wave spin-singlet pairing with a triangular-lattice $t$$-$$J$ 
model~\cite{Baskaran03,Kumar03,Wang04,Ogata03} and the {\it f}-wave spin-triplet 
pairing with extended Hubbard model~\cite{Ikeda04,Tanaka04-2,Kuroki04-1} 
have been proposed. Also, the {\it p}-wave spin-triplet has been suggested 
phenomenologically~\cite{Tanaka03}.

The third candidate is Sr$_2$RuO$_4$, which is a tetragonal, layered perovskite 
system of stacking RuO$_2$-planes. No sooner was discovered the superconducting 
transition at $T_{\rm c} \sim 1.5$ K~\cite{Maeno94}, the possibility of spin-triplet 
pairing was pointed out~\cite{Rice95}. After that, the spin-triplet 
pairing state with {\bf d}-vector perpendicular to the conducting plane has been 
confirmed by a NMR measurement~\cite{Ishida98}. Like Cu in the high-$T_{\rm c}$ 
superconductors, Ru forms a square lattice. If a single-band description of 
RuO$_2$-plane for so-called $\gamma$ band could be adequate, the system is 
described as a 2D Hubbard model with next-nearest-neighbor hopping. 
The next-nearest-neighbor hopping has been estimated to be $0.3 - 0.4$ in units 
of the nearest-neighbor hopping~\cite{Mazin97,Liebsch00}, so that Ru indeed 
forms a triangular network. Based on the quantum  oscillation  measurement, 
the $\gamma$ Fermi surface sheet is a large electron-like cylinder with the electron 
filling $n \sim 2/3$. At present, there exist several theoretical studies where 
a competition between {\it d}-wave spin-singlet and {\it f}-wave spin-triplet 
pairing states is discussed~\cite{Hasegawa00,Nomura00,Kuwabara00,Kuroki04-2}.

In our previous works~\cite{Ohta05,Nishimoto08}, we studied the ground-state 
properties of two-chain Hubbard model coupled with zigzag bonds. This system 
is equivalent to a 1D triangular-lattice Hubbard model. We argued there that 
a cyclic hopping  motion of two electrons in each triangle yields a ferromagnetic 
correlation in the strong-coupling regime; as a consequence, a spin-triplet 
superconducting state becomes dominant if the following conditions are satisfied: 
(i) The product of three hopping integrals in each triangle is positive. 
(ii) The zigzag bond is rather weaker than the intra-chain bond. 
(iii) The filling is substantially away from zero and half fillings. 
The three-site cyclic-hopping mechanism should work even in 2D system, 
and it is particularly worth noting that the above-mentioned three materials 
completely meet the conditions. We therefore consider a 2D anisotropic 
triangular-lattice Hubbard model. We employ the density-matrix renormalization 
group (DMRG) method~\cite{White92} to calculate the spin-spin correlation 
and spin-triplet pair-correlation functions. It will thereby be confirmed 
that the spin-triplet pairing occurs predominantly in the {\it f}-wave 
channel for a simple 2D Hubbard model.

\begin{figure}[t]
\includegraphics[width= 7.0cm,clip]{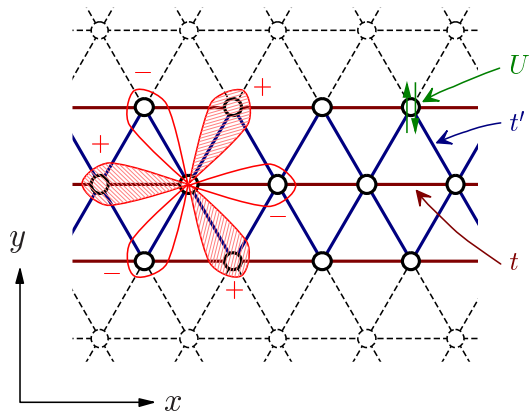}
\caption{(Color online) Schematic representation of the anisotropic triangular-lattice 
Hubbard model. Lattice drawn with bold lines is extracted for our calculation 
of the spin-triplet pair-correlation functions. The f-wave pair function is also 
shown.
}
\label{lattice}
\end{figure}

The Hamiltonian of the anisotropic triangular-lattice Hubbard model is given by
\begin{equation}
\nonumber
H=\sum_{\left\langle ij\right\rangle \sigma}t_{ij}
(c_{i\sigma}^\dag c_{j\sigma}+{\rm H.c.}) 
+U\sum_i n_{i\uparrow}n_{i\downarrow}
\label{hamiltonian}
\end{equation}
where $c_{i\sigma}^\dag$ ($c_{i\sigma}$) creates (annihilates) 
an electron with spin $\sigma$ at site $i$,  
$n_{i\sigma}=c_{i\sigma}^\dag c_{i\sigma}$ is a number operator, 
and $t_{ij}$ is hopping integral between sites $i$ and $j$. 
The sum $\left\langle ij \right\rangle$ runs over nearest-neighbor 
pairs. We here include two kinds of nearest-neighbor hopping integrals 
$t$ in the $x$-direction and $t^\prime$ otherwise, as shown in 
Fig.~\ref{lattice}. Onsite Coulomb repulsion $U$ is set to 
be $10t$ as a typical value in the strongly-correlated electrons system. 
We take $t=1$ as the energy unit hereafter.

\begin{figure}[b]
\includegraphics[width= 6.5cm,clip]{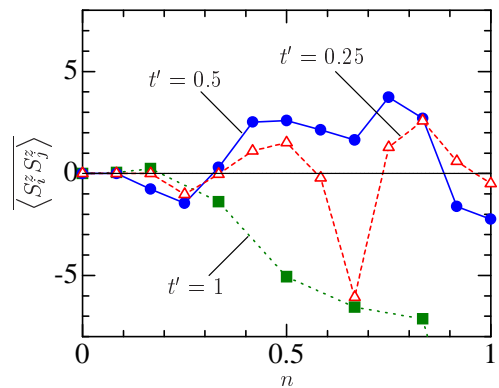}
\caption{(Color online) Average spin-spin correlations 
$\overline{\left\langle S_i^z S_j^z \right\rangle}$ between sites 
bonded with $t^\prime$, as a function of filling $n$.
}
\label{fig2}
\end{figure}

Let us first demonstrate that the cyclic-hopping mechanism indeed 
leads to ferromagnetic correlation in the 2D system (\ref{hamiltonian}). 
We thus study a finite-size cluster of $l_x=6$ and $l_y=8$, where 
the periodic (open) boundary conditions are applied for the $x$- 
($y$-) direction. This choice of the boundary conditions enables us 
to carry out sufficiently accurate calculations. We keep up to $m=3400$ 
density-matrix eigenstates in the DMRG procedure. In this way, 
the maximum truncation  error, i.e., the discarded weight, is less than  
$1 \times 10^{-5}$ and the maximum error in the ground-state energy 
is estimated to be $E_0 /(l_xl_y) \sim 10^{-2}$. Figure~\ref{fig2} shows 
the average spin-spin correlations $\overline{\left\langle S_i^z S_j^z \right\rangle}$ 
between the neighboring sites bonded with $t^\prime$, where 
$\left\langle \cdots \right\rangle$ denotes the ground-state expectation value. 
We can see the enhancement of ferromagnetic correlation in a wide 
range of filling for $t^\prime=0.25$ and $0.5$. The suppression 
of the correlation around $n=0.7$ implies that the origin of 
ferromagnetism for $n \gtrsim 0.7$ is different from that for 
$n \lesssim 0.7$: it comes from the cyclic-hopping mechanism around 
$n=0.5$ and from the Nagaoka-Thouless mechanism near $n=1$.
Those behaviors are qualitatively consistent with the results in the 1D 
triangular-lattice Hubbard model~\cite{Nishimoto08}. On the other hand, 
antiferromagnetic correlation is mostly dominant for $t^\prime=1$ 
since the direct exchange interaction will exceed the ferromagnetic ones. 
We note that the correlation seems to be slightly ferromagnetic for 
$n \lesssim 0.2$, where the nearly flat-band system may be realized. 
So we confirm that in the 2D triangular-lattice Hubbard model the ferromagnetic 
correlation induced by the cyclic-hopping mechanism can occur at $n \sim 0.5$ 
if the anisotropy is rather strong.

\begin{figure}[t]
\includegraphics[width= 6.5cm,clip]{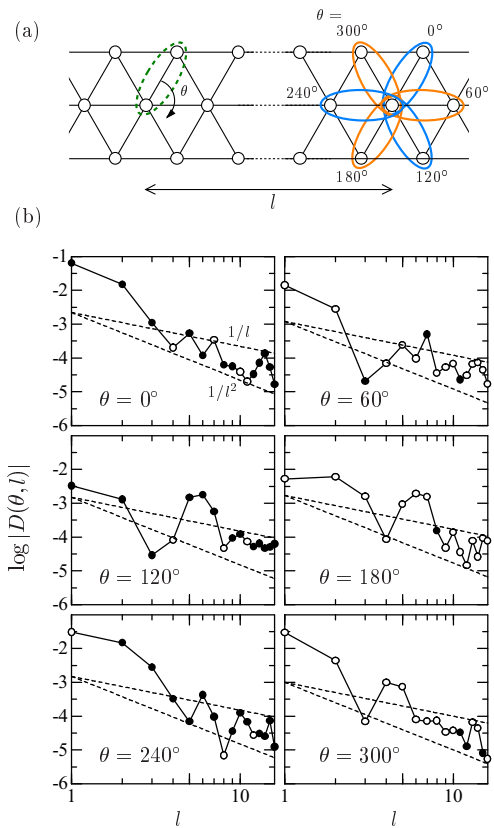}
\caption{(Color online) (a) Definition of the rotation angle $\theta$ and 
the distance $l$. For a case that the pair is removed from a position $i$ 
with dashed ellipse, the position of the pair created at $i+l$ is denoted 
by solid ellipse for each $\theta$. (b) $\log |D(\theta,l)|$ with $l$ at 
$t^\prime=0.5$ for all possible $\theta$. Filled (empty) symbol denotes 
a positive (negative) value of $D(\theta,l)$.
}
\label{fig3}
\end{figure}

We next turn to the long-range behavior and the spatial symmetry of 
spin-triplet pair-correlation functions
\begin{eqnarray}
D(\theta,l)=\left\langle \Delta_{i+l}^\dagger \Delta_i \right\rangle
\label{Dl}
\end{eqnarray}
with a spin-triplet operator 
$\Delta_i=c_{i\uparrow}c_{i+r\downarrow}+c_{i\downarrow}c_{i+r\uparrow}$ 
where $i+r$ denotes the neighboring site of $i$. As in Fig.~\ref{lattice} 
we extract a three-leg ladder from 2D lattice, which is a minimal model to 
generate all possible pairing symmetries. In Eq.(\ref{Dl}), we 
remove a pair from sites $(i,i+r)$ and add the pair at sites 
$(i+l,i+l+r^\prime)$, whereby we rotate the pair by $\theta$ degree from 
$i \to r$ direction to $i \to r^\prime$ clockwise [see Fig.~\ref{fig3}(a)]. 
Note that the same results must be expected even if we define the spin-triplet 
operator as $\Delta_i=c_{i\uparrow}c_{i+r\uparrow}$ or 
$c_{i\downarrow}c_{i+r\downarrow}$. We also restrict ourselves to the case 
at quarter filling, i.e., $n = 0.5$. We calculate the pair-correlation functions 
(\ref{Dl}) of the three-leg ladder with the DMRG method. The OBC are 
applied in the {\it x}-direction so that the correlation functions will be 
calculated using distances taken about the midpoint. In the following, 
we study a ladder with $32 \times 3$ sites with keeping $m=4500$ density-matrix 
eigenstates to build the DMRG basis. The obtained ground-state energy is 
expected to be accurate to parts in $10^{-2}t$. In Fig.~\ref{fig3}(b), 
the pair-correlation functions $D(\theta,l)$ are plotted as a function of 
distance $l$ for all possible $\theta$ values with a fixed anisotropy $t^\prime=0.5$. 
We can see that $D(\theta,l)$ decay as $\sim 1/l^\gamma$ $(1<\gamma<2)$ 
for all $\theta$ values. The exponent $\gamma$ smaller than $2$ would imply 
an attractive interaction between electrons~\cite{Schulz96}. It is consistent 
with our previous results~\cite{Ohta05}. We also find that the values of 
$D(\theta,l)$ are mostly positive for $\theta=0^\circ, 120^\circ, 240^\circ$ 
and, whereas, negative for $\theta=60^\circ, 180^\circ, 300^\circ$. 
In other words, the pair wave function changes its sign by $\pi/3$ rotation.

\begin{figure}[t]
\includegraphics[width= 7.0cm,clip]{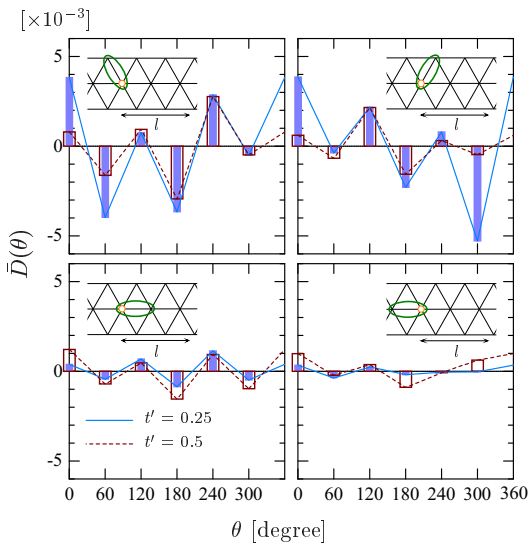}
\caption{(Color online) Rotation-angle dependence of the sum of 
the spin-triplet pair-correlation functions $\bar{D}(\theta)$ for $J=J^\prime=0$. 
Inset shows the position of the pair annihilated by the pair operator 
$\Delta_i$.
}
\label{fig4}
\end{figure}

In order to organize the angle dependence of $D(\theta,l)$ more explicitly, 
we sum it up for pair separations $l=7-25$: 
\begin{eqnarray}
\bar{D}(\theta)=\sum_{l=7}^{25}D(\theta,l).
\label{D}
\end{eqnarray}
The open-end effects are expected to be rather small for central $25$ sites of 
the system ($l \lesssim 25$). In Fig.\ref{fig4}, the results of $\bar{D}(\theta)$ 
with $t^\prime=0.25$ and $0.5$ are shown. For both parameters we can see 
that the sign of $\bar{D}(\theta)$ changes by $\pi/3$ rotation, which clearly 
indicates the {\it f}-wave spatial symmetry of the pair correlation function. 
Since the spin-triplet pairs are formed mostly on the zigzag bond for 
$t^\prime=0.25$~\cite{Ohta05}, $\bar{D}(\theta)$ involving the on-leg pairs, namely, 
at $\theta=2\pi/3, 5\pi/3$ in the upper left panel, $\theta=\pi/3, 4\pi/3$ in the upper 
right panel, and all of $\theta$ in the lower panels, are relatively small. 
When $t^\prime$ is increased from $0.25$ to $0.5$, the pairing correlation 
on the zigzag bond is somewhat reduced and, whereas, that on the leg bond 
is enhanced. The situation is rather complicated: $t^\prime$ enhances 
the antiferromagnetic correlation on the zigzag bond by the direct exchange 
interaction (scaled as $\propto t^{\prime2}$), and also enhances the ferromagnetic 
correlation on each triangle by the three-site cyclic-hopping exchange interaction 
(scaled as $\propto tt^{\prime2}$). Consequently, the {\it f}-wave symmetry 
of the superconducting state becomes more isotropic. However, we note that 
the spin-triplet pair correlation becomes smaller as the system approaches to 
an isotropic triangle lattice. 

In summary, the ground state of the 2D triangular-lattice Hubbard model is studied 
with the DMRG method. We find that the three-site cyclic-hopping mechanism 
really induces ferromagnetic correlation around quarter filling, which is 
mandatory for the spin-triplet pairing, provided that the anisotropy is rather 
strong. Slow decay of the pair-correlation functions with distance is indicative 
of the dominant spin-triplet superconductivity. The pair-correlation function 
changes its sign by $\pi/3$ rotation, indicating the {\it f}-wave rotational 
symmetry. Thus, we suggest a new mechanism of the {\it f}-wave spin-triplet 
superconductivity derived from the three-site cyclic-hopping ferromagnetic interactions. 
We also point out that the mechanism may be possibly relevant to the spin-triplet 
superconducting systems, such as Bechgaard salts (TMTSF)$_2$X, cobalt oxide 
Na$_{0.35}$CoO$_2$$\cdot$$1.3$H$_2$O, and layered perovskite Sr$_2$RuO$_4$.

\acknowledgments

This work was supported in part by Grants-in-Aid for Scientific Research
(Nos. 18540338, 18028008, 18043006, and 19014004) from the Ministry of 
Education, Science, Sports, and Culture of Japan.  A part of computations 
was carried out at the Research Center for Computational Science, Okazaki 
Research Facilities, and the Institute for Solid State Physics, 
University of Tokyo.

\end{document}